\documentclass[12pt]{iopart}

\newcommand{\be}{\begin{eqnarray}}
\newcommand{\ee}{\end{eqnarray}}
\newcommand{\bea}{\begin{eqnarray}}
\newcommand{\eea}{\end{eqnarray}}

\usepackage{graphicx}
\usepackage{bm}
\usepackage{color}
\usepackage{slashed}
\usepackage{cite}
\usepackage{graphicx}
\usepackage{multicol}
\usepackage{graphicx}
\usepackage{color}
\usepackage{slashed}
\usepackage{CJKutf8}
\begin{document}
\begin{CJK}{UTF8}{<font>}

\title{Cooling-heating phase transition and critical behavior of the charged accelerating AdS black hole}

\author{Sen Guo$^{1}$, \ Ya-Ling Huang$^{2}$, \ Guo-Ping Li$^{3*}$}

\address{
$^1$Guangxi Key Laboratory for Relativistic Astrophysics, School of Physical Science and Technology, Guangxi University, Nanning 530004, People's Republic of China\\
$^2$School of Electrical Engineering, SouthWest JiaoTong University, Chengdu 610031, People's Republic of China\\
$^3$School of Physics and Astronomy, China West Normal University, Nanchong 637000, People's Republic of China}

\ead{sguophys@126.com; katrina996@163.com; gpliphys@yeah.net}
\vspace{10pt}
\begin{indented}
\item[]Oct 2021
\end{indented}

\begin{abstract}
We study the cooling-heating phase transition of the charged accelerating anti-de Sitter black hole in extended phase space, and investigate the critical behavior of this black hole in extended phase space. By calculating the thermodynamic quantities and state equation, we found that the charged accelerating AdS black hole as thermodynamic system is similar to the van der Waals system. The inversion temperature of this black hole is obtained, and cooling-heating and isenthalpic curves are plotted in $T-P$ plane. Our results indicate that the inversion temperature for a given pressure increases with $e$, and the acceleration parameter has the opposite effect, which the cooling-heating curves decreases gradually with the the increases of $a$. We also analyse the influence of acceleration parameter on isenthalpic curves, implying that the phase transition point decreases with the increase of acceleration factor under constant pressure.
\end{abstract}


\section{Introduction}
\label{intro}
\par
Black hole thermodynamics can be used as a bridge connect classical thermodynamics, quantum mechanics and general relativity, which can be consider as a topic of great concern in theoretical physics. The various of the black holes thermodynamic properties have been investigated since Hawking radiation and Bekenstein-Hawking entropy were first proposed \cite{1,2,3,4,5,6}. When the black hole as a thermodynamic system, it shows that some interest thermodynamic properties similar to the classical thermodynamic system. By studying the phase transition of the Schwarzschild-AdS black hole, Hawking and Page found that these similarities become more accurate and obvious in anti-de Sitter (AdS) space-time \cite{7}. Furthermore, the research shown that the phase transition of the Reissner-Nordstr\"{o}m-AdS black hole can be compared to the van der Waals fluid system \cite{8}.

\par
Recently, the idea of the cosmological constant as thermodynamic pressure in extended phase space leads the research direction, i.e. $P=-\Lambda/8\pi=3/8 \pi l^{2}$ \cite{9,10}. By analyzing the charged AdS black hole phase transition, Gunasekaran $et.al$ found that the charged AdS black hole have similar $P-\upsilon$ diagram and critical exponents \cite{11}. Subsequently, this pioneering work has been generalized to other kinds of black holes \cite{12,13,14,15,16,17,18,19,20,21,22,23,24}. Apart from the thermodynamic phase transition and critical phenomena, the another important study of black hole thermodynamics is cooling-heating phase transition. \"{O}kc\"{u} and Ayd{\i}ner creatively extends the famous Joule-Thomson expansion process to black hole system, showing that the Reissner-Nordstr\"{o}m-AdS black hole exist inversion temperature and inversion curves, and also have cooling-heating regions. Subsequently, the Joule-Thomson expansion of various black holes has been investigated, such as d-dimensional charged AdS black hole \cite{26}, Gauss-Bonnet black holes \cite{27}, regular(Bardeen)-AdS black hole \cite{28}, $f(R)$ gravity coupled with Yang-Mills field \cite{29} and Rastall gravity \cite{29}, etc. As a result, these researches suggest that the different gravities backgrounds have effects on cooling-heating phase transition process.

\par
We known that an interest kind of black hole solution, which takes the form of a cone defect angle angle attached to the polar axis of the black hole. Because of this defect, it provides the driving force for acceleration, which can remove accelerated horizon, this kind of black hole is called accelerating black hole. It is thought to be due to the creation of a force that moves it away from the center of the negatively curved space-time, and the cosmic string terminates at the event horizon \cite{30}. In \cite{31,32}, C-metric represents the metric of the black hole, but this solution is idealized, because the singularity cone of a black hole can be replaced by a finite width cosmic string core \cite{33}. Meanwhile, the application of C-metric is not limited to general relativity, which can be used to describe the generation of black holes in an electric or magnetic field and the splitting of cosmic strings \cite{33,34,35}. However, people found that the research of its thermodynamic properties is somewhat abstruse, which is due to unusual asymptotic properties and at least one immovable conic singularity in symmetric azimuth axis \cite{36,37,38,39,40}.

\par
Nevertheless, the cooling-heating phase transition and critical behavior of the charged accelerating AdS black hole is still of opening question. This paper focuses on this issue. We calculate the equation of the state of this black hole and investigate the $P-\upsilon$ critical behavior. By considering the cooling-heating phase transition, we derive the inversion temperature of this black hole and plot the inversion and isenthalpic curves. We also analyse the influence of acceleration parameter and black hole charge on thermodynamic behaviors. The outlines of this paper are listed as follows. In Sec. \ref{sec:2}, we mainly review the thermodynamics properties of the charged accelerating AdS black hole in extended phase space, and investigate to critical behavior of this black. Section \ref{sec:3} discussed the cooling-heating phase transition of this black hole. The summary and discussion for this paper is presented in Sec. \ref{sec:4}.

\section{The thermodynamic properties of the charged accelerating AdS black hole and $P-\upsilon$ critical}
\label{sec:2}
\par
The line element of this black hole can be expressed as \cite{30}
\begin{equation}
\label{2-1}
ds^{2}=\frac{1}{\Omega^{2}}\Big[f(r)dt-\frac{d r^{2}}{f(r)}-r^{2}\Big(\frac{d\theta^{2}}{g(\theta)}+g(\theta)\sin^{2}\theta \frac{d\phi^{2}}{K^{2}}\Big)\Big],
\end{equation}
in which the $\Omega$ is the conformal factor,
\begin{equation}
\label{2-2}
\Omega=1+A r \cos \theta,
\end{equation}
the existence of conformal factor can be used to guarantee the conformal invariance and boundary conditions in AdS spacetime. The $f(r)$ is
\begin{equation}
\label{2-3}
f(r)=(1-A^{2}r^{2})\Big(1-\frac{2m}{r}+\frac{e^{2}}{r^{2}}+\frac{r^{2}}{l^{2}}\Big),
\end{equation}
where $A$ is acceleration parameter, $M$ is the mass and $e$ is the charge of the black hole. The g($\theta$) is
\begin{equation}
\label{2-4}
g({\theta})=1+2 m A {\cos}{\theta}+e^{2}A^{2}{\cos}^{2}{\theta}.
\end{equation}
The black hole have the following regularity at the pole
\begin{equation}
\label{2-5}
K_{\pm}=g(\theta_{\pm})=1 \pm 2 A m + e^{2} A^{2}.
\end{equation}
From the above formula, we can see that the positive and negative of the parameter $K$ must be fixed, otherwise both solutions are regular. One can obtain that
\begin{equation}
\label{2-6}
K=1 + 2 A m + e^{2} A^{2}.
\end{equation}
By integrating on the conformal infinite sphere, the black hole mass, charge and potential can be written as
\begin{equation}
\label{2-7}
M=\frac{m}{K},~~~~Q=\frac{1}{4 \pi}\int_{\Omega=0}F=\frac{e}{K},~~~~\Phi=\frac{e}{r_{h}},
\end{equation}
where $r_{h}$ is the event horizon radius of this black hole, $F$ is electromagnetic field tensor, depending on the standard potential $B$,
\begin{eqnarray}
\label{2-8}
B=-\frac{e}{r}dt,~~~F=dB.
\end{eqnarray}
The area of the horizon as
\begin{equation}
\label{2-9}
A=\int_{0}^{\pi}\int_{0}^{2\pi}\sqrt{g_{\theta \theta}g_{\phi\phi}}d\theta d\phi =\frac{4 \pi r_{h}^{2}}{K(1-A^{2}r_{h}^{2})}.
\end{equation}
Hence, the black hole entropy can be written as
\begin{equation}
\label{2-10}
S=\frac{A}{4}=\frac{\pi r_{h}^{2}}{K(1-A^{2}r_{h}^{2})}.
\end{equation}
According to equation (\ref{2-3}), the black hole mass is given by
\begin{equation}
\label{2-11}
m=\frac{3r_{h}^{2}+(8\pi P - 3A^{2})r_{h}^{4}+3e^{2}(1-A^{2}r_{h}^{2})}{6 r_{h}(1-A^{2}r_{h}^{2})},
\end{equation}
and the black hole temperature is
\begin{equation}
\label{2-12}
T=\frac{(1-A^2 r_{h}^{2})(8P \pi r_{h}^{4}+r_{h}^{2}-e^2)}{4 \pi r_{h}^{3}}.
\end{equation}
The pressure expression is obtained according to equation (\ref{2-12})
\begin{equation}
\label{2-13}
P=\frac{A^2 e^2 r_{h}^{2}+r_{h}^{2}-e^2-A^2 r_{h}^{4}-4 \pi r_{h}^{3} T}{8 \pi r_{h}^{4} (A^2 r_{h}^{2}-1)},
\end{equation}
where the $A$ is acceleration term, satisfying $A=a/r_{h}$ \cite{30}. The equation of state for this black hole is obtained, we have
\begin{equation}
\label{2-14}
P=\frac{(a^2 -1)e^2- r_{h}^{2}(a^{2}-1+4 \pi r_{h} T)}{8 \pi r_{h}^{4} (a^2-1)}.
\end{equation}
Moreover, we can get several other thermodynamic variables according to the first law of thermodynamics,
\begin{eqnarray}
\label{2-15}
V=\Big(\frac{\partial {M}}{\partial P}\Big)_{S,Q},~~\Phi=\Big(\frac{\partial {M}}{\partial Q}\Big)_{S,P}.
\end{eqnarray}
At the critical point, we have \cite{11}
\begin{equation}
\label{2-16}
\frac{\partial P}{\partial r_{h}}=\frac{{\partial}^{2}P}{\partial r_{h}^{2}}=0.
\end{equation}
According to equations (\ref{2-14}) and (\ref{2-16}), the black hole critical physical quantity as
\begin{eqnarray}
\label{2-17}
r_{c}=\sqrt{6}e,~~~P_{c}=\frac{{(a^{2}-1)}^{2}}{32(3-a^{2})e^{2}\pi},~~~T_{c}=\frac{1-a^{2}}{3\sqrt{6}e \pi}.
\end{eqnarray}
The universal ratio can be calculated, i.e.
\begin{equation}
\label{2-18}
\epsilon=\frac{P_{c}\nu}{T_{c}}=\frac{9(a^{2}-1)}{8(a^{2}-3)},
\end{equation}
in which the specific volume $\nu$ is twice the size of $r_{h}$ \cite{12}. However, this result is different from the van der Waals system ($3/8$). Note that we taken the acceleration parameter $a$ as zero, the equation (\ref{2-18}) can be rewritten as
\begin{equation}
\label{2-19}
\epsilon=\frac{P_{c}\nu}{T_{c}}=\frac{3}{8}.
\end{equation}
\begin{center}
\includegraphics[width=7cm,height=5cm]{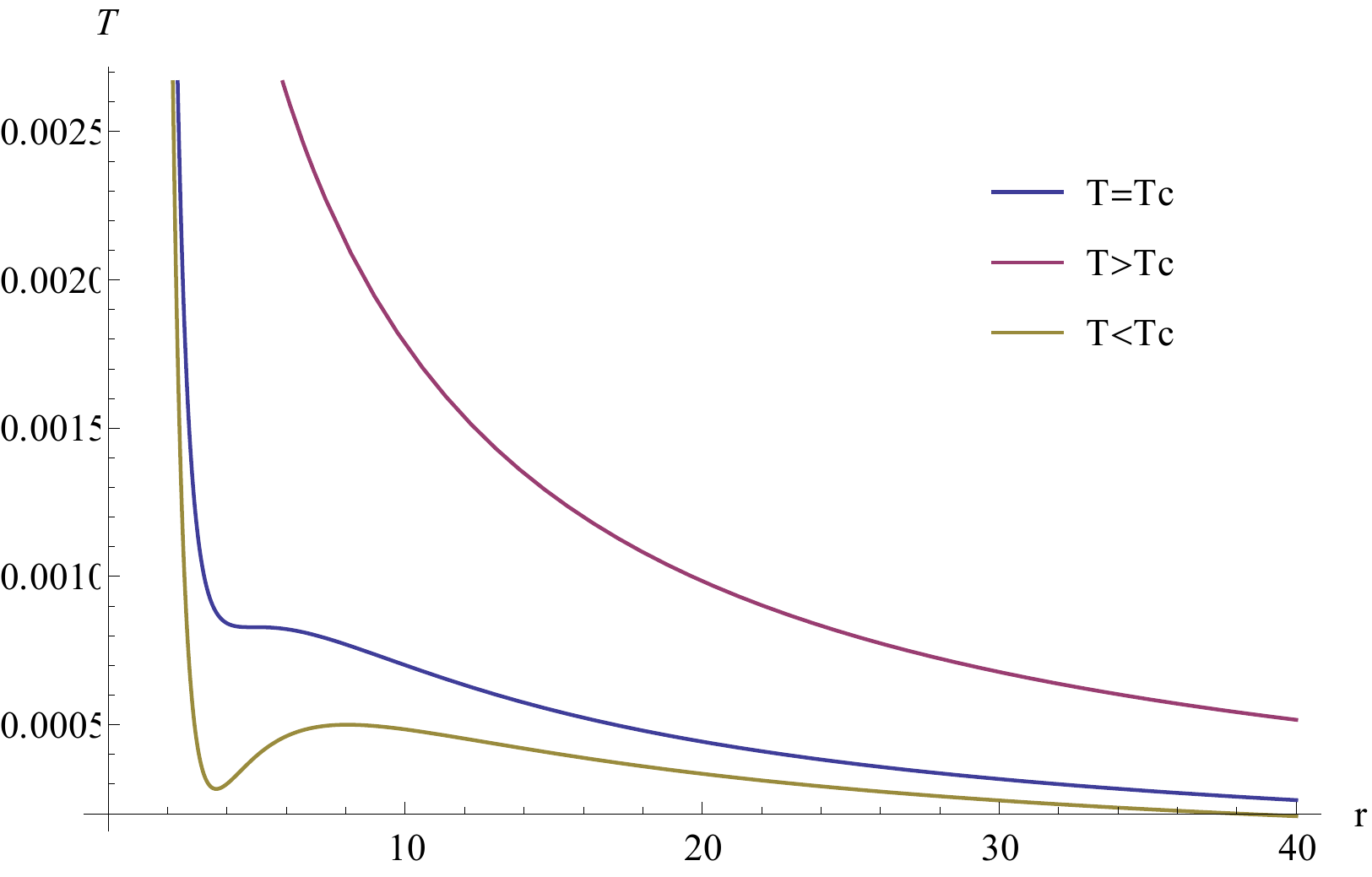}
\includegraphics[width=7cm,height=5cm]{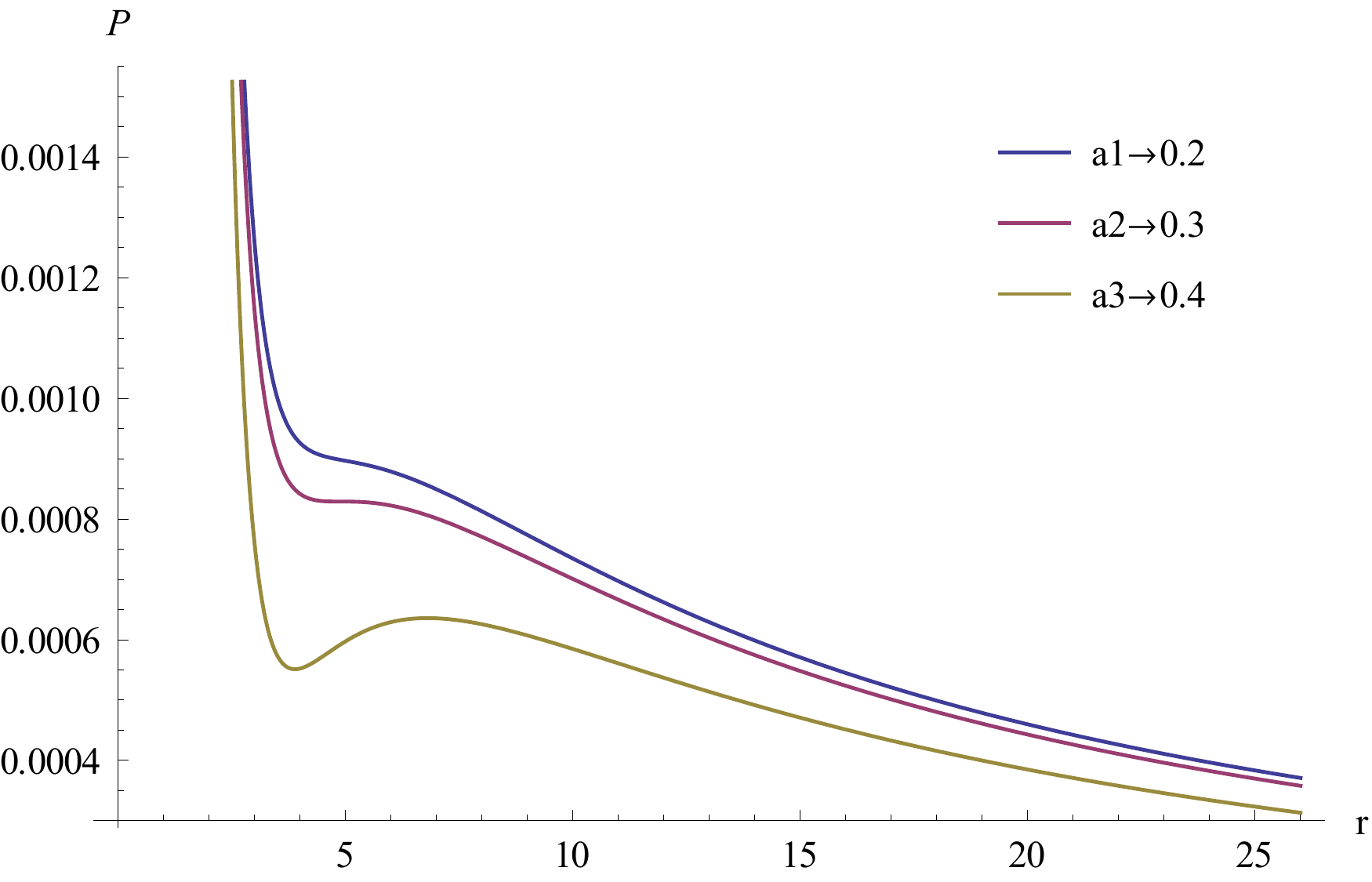}
\parbox[c]{15.0cm}{\footnotesize{\bf Fig~1.}  
$P-r$ diagram of the charged accelerating AdS black hole. The red line, blue line and green line correspond to $T>T_{c}$, $T=T_{c}$ and $T<T_{c}$ in the left panel, respectively. The right panel represent $a=0.2,0.3,0.4$ from top to bottom. Taking $a=0.2$ and charge as $e=2$.}
\label{fig:1}
\end{center}
Figure 1 upper curve ($T>T_{c}$) shows that the single phase properties of ideal gas, meaning the black hole no phase transition and represents only stable state. The lower curve ($T<T_{c}$) correspond to unstable state, which also behalf of the phase of gas-liquid coexistence. The small/lagre black hole will be phase transition at $T<T_{c}$, and the black hole no phase transition at no phase transition. The phase transition curves move to the down gradually with the increase of the acceleration parameter $a$ when the charge $e$ is constant, which indicating that the phase transition point decreases with the increase of $a$.

\par
Meanwhile, the critical exponents come from ref. \cite{13}
\begin{eqnarray}
\label{2-20}
&C_{v}=\Big(T\frac{\partial S}{\partial T}\Big)_{V}\propto \mid t \mid^{-\alpha},~~~\eta=V_{l}-V_{s}\propto \mid t \mid^{\beta},\nonumber\\
&\kappa_{T}=\Big(-\frac{1}{V}\frac{\partial V}{\partial P}\Big)_{T} \propto \mid t \mid^{-\gamma},~~~(P-P_{c}) \propto (V-V_{c})^{\delta}.
\end{eqnarray}
By solving equations (\ref{2-14}) and (\ref{2-20}), the phase transition critical exponents of the charged accelerating AdS black hole is $(\alpha,\beta,\gamma,\delta)$ =(0,~1/2,~1,~3). As a result, it is similarities with van der Waals system and Reissner-Nordstr\"{o}m-AdS black hole \cite{15}.

\section{Cooling-heating phase transition of the charged acclerating AdS black holes}
\label{sec:3}
\par
The Joule-Thomson expansion is famous classical thermodynamic process, which shown that the gas undergoes irreversible adiabatic expansion from high pressure to low pressure through a porous plug or valve, and the most obvious feature of the process is that the enthalpy remains unchanged. The rate of change of gas temperature with pressure is defined as cooling-heating coefficient \cite{25}
\begin{equation}
\label{3-1}
\mu={\Big(\frac{\partial T}{\partial P}\Big)}_{H}.
\end{equation}
When the gas expands, the pressure is always lower and always shows negative, hence, one can consider that the sign of cooling-heating coefficient can determine the cooling and heating of the system. Concretely, the temperature is always positive when $\mu<0$, corresponding to the warm system, and vice versa. According to \cite{25}, the cooling-heating coefficient can be written as
\begin{equation}
\label{3-2}
\mu={\Big(\frac{\partial T}{\partial P}\Big)}_{H}=\frac{1}{C_p}\Big[T \Big({\frac{\partial V}{\partial T}}\Big)_p-V\Big].
\end{equation}
Setting $\mu=0$, the inversion temperature is given by
\begin{equation}
\label{3-3}
T_{i}=V {\Big(\frac{\partial T}{\partial V}\Big)}_{P}.
\end{equation}

\par
Based on these results, the cooling-heating phase transition of the charged accelerating AdS black hole have been investigated in this section. We known that the black hole enthalpy is determined as the black hole mass in extended phase space, hence the black hole enthalpy remains constant during cooling-heating phase transition process. Therefore, the pressure $P$ can be written as a function of the black hole mass $m$ and radius $r_{h}$ according to equations (\ref{2-3}) and (\ref{2-13}), we have
\begin{equation}
\label{3-4}
P(m,r_{h})=\frac{6 m r_{h}-3 e^{2}-3 r_{h}^{2}}{8 \pi r_{h}^{4}},
\end{equation}
and substituting upper formula into equation (\ref{2-12}), one can obtain
\begin{equation}
\label{3-5}
T(m,r_{h})=\frac{(a^2 -1)(2e^{2}-3 m r_{h}+r_{h}^{2})}{2 \pi r_{h}^{3}}.
\end{equation}
From the relation between the cooling-heating coefficient, equation (\ref{2-19}) can be re-written as
\begin{equation}
\label{3-6}
\mu={\Big(\frac{\partial T}{\partial P}\Big)}_{H}={\Big(\frac{\partial T}{\partial {r_{h}}}\Big)}_{H}{\Big(\frac{\partial {r_{h}}}{\partial P}\Big)}_{H}=\frac{\Big(\partial T/\partial {r_{h}\Big)}_{H}}{\Big(\partial P/\partial {r_{h}}\Big)_{H}}.
\end{equation}
According to equations (\ref{3-4}), (\ref{3-5}) and (\ref{3-6}), the cooling-heating coefficient as
\begin{equation}
\label{3-7}
\mu=\frac{4 r_{h} (a^{2}-1) (2 r_{h}^{2}- 3e^{2}+ 8 \pi P r_{h}^{4})}{3(e^{2}-r_{h}^{2}-8 \pi P r_{h}^{4})}.
\end{equation}
Setting $\mu=0$, one can get
\begin{equation}
\label{3-8}
2 r_{h}^{2}- 3e^{2}+ 8 \pi P r_{h}^{4}=0.
\end{equation}
By solving above equation for $r_{h}(P_{i})$, we choose a real number solution with physical meaning, i.e.
\begin{equation}
\label{3-9}
r=\frac{\sqrt{\frac{\sqrt{1+24e^{2}P_{i} \pi}}{P_{i} \pi}-\frac{1}{P_{i} \pi}}}{2\sqrt{2}},
\end{equation}
where the $P_{i}$ is inversion pressure. Substituting this root into equation (\ref{2-12}), the inversion temperature is obtained, we have
\begin{equation}
\label{3-10}
T_{i}=\frac{(a^{2}-1)(\sqrt{1+24 P_{i}e^{2}\pi}-1-16 e^{2}P_{i}\pi)}{P_{i}\sqrt{2 \pi}\Big(\frac{\sqrt{1+24e^2 P_{i}\pi}-1}{P_{i}}\Big)^{3/2}}.
\end{equation}

\par
According to the expression of equation (\ref{3-10}), the charged accelerating AdS black hole inversion temperature curves have been plotted in Figure 2. One can see that the increase of $e$ leads to the increase of inversion temperature at the pressure constant, implying that the higher the temperature is needed to complete the cooling-heating phase transition for the larger charged the black hole. Note that the values of $a$ is sensitive to inversion temperature. The inversion temperature curve is lower for a larger $a$.
\begin{center}
\includegraphics[width=7cm,height=6cm]{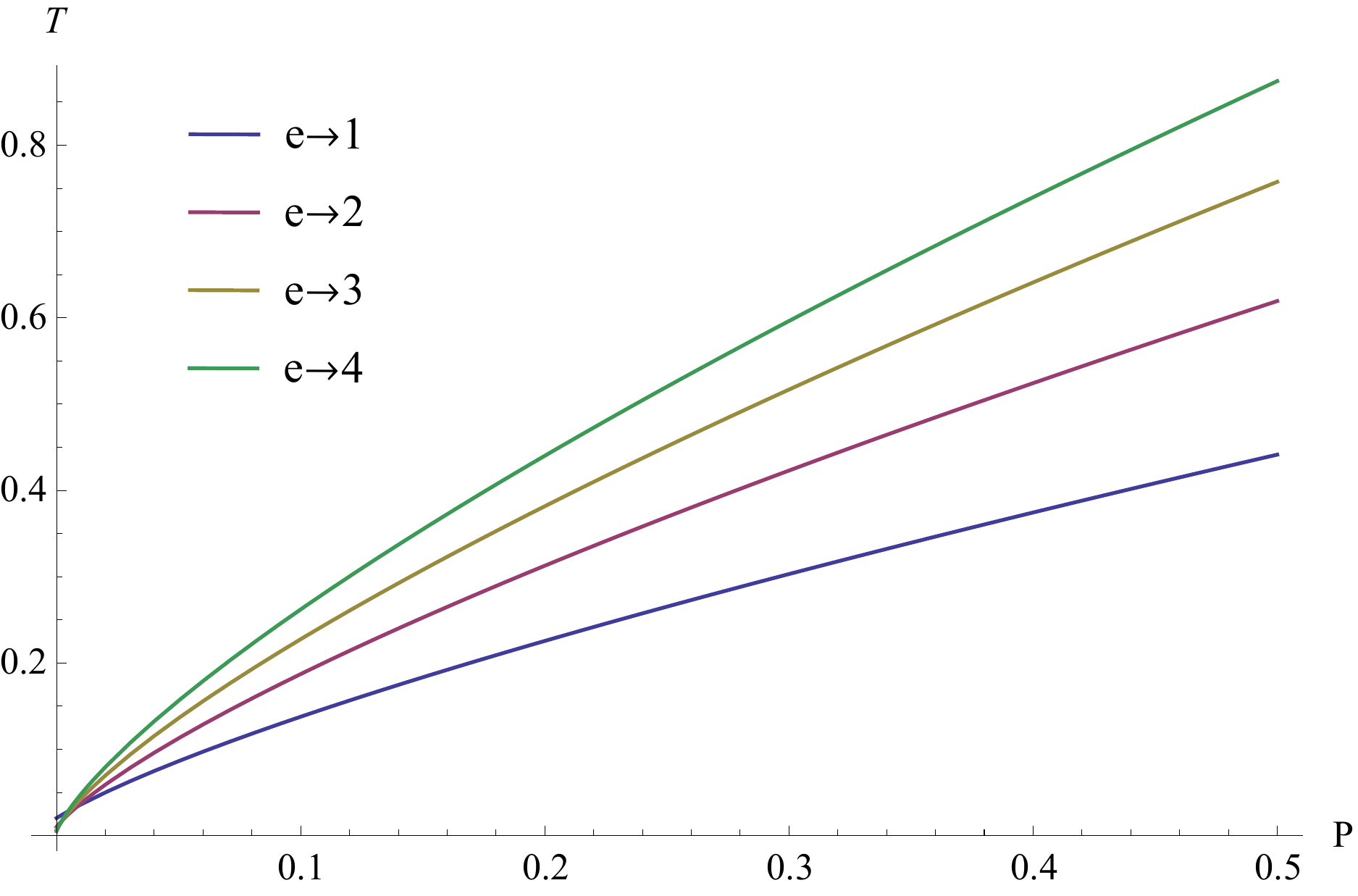}
\includegraphics[width=7cm,height=6cm]{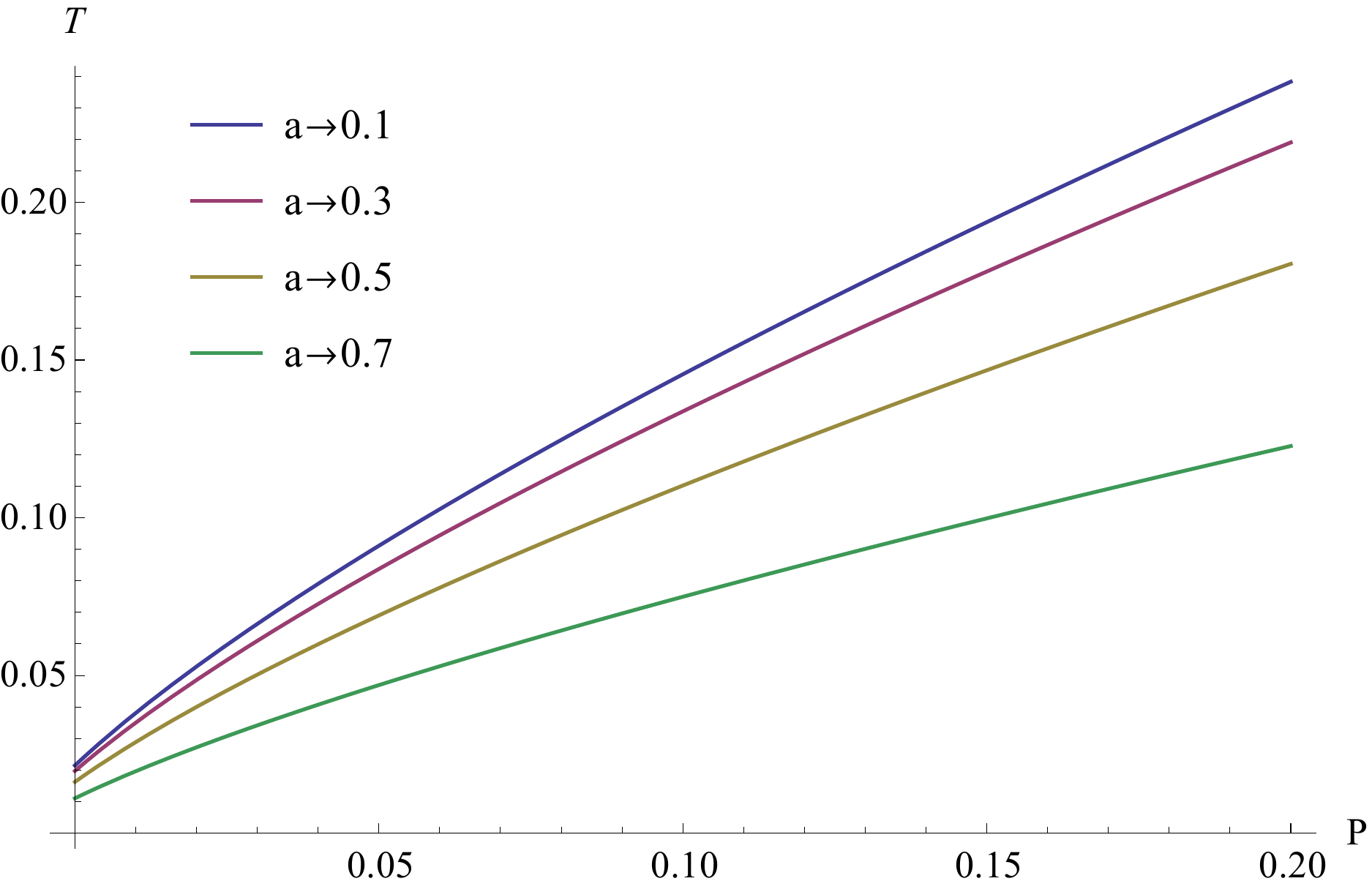}
\parbox[c]{15.0cm}{\footnotesize{\bf Fig~2.}  
Inversion curves of the charged accelerating AdS black hole. {\em Panel (a)}-- black hole charge as $e=1,2,3,4$ and accelerating parameter as $a=0.2$, {\em Panel (b)}-- accelerating parameter as $a=0.7,0.5,0.3,0.1$ and black hole charge as $e=1$.}
\label{fig:2}
\end{center}
When the inversion pressure $P_{i}$ is zero, one can get the minimum inversion temperature,
\begin{equation}
\label{3-11}
T_{i}^{min}=\frac{1-a^{2}}{6\sqrt{6}\pi e}.
\end{equation}
The ratio between the minimum inversion temperature and the critical temperature is obtained, we have
\begin{equation}
\label{3-12}
\frac{T_{i}^{min}}{T_{c}}=\frac{1}{2}.
\end{equation}
Note that this ratio of the charged accelerating AdS black hole is smaller than that of the van der Waals fluid (0.75). However, this result is the same as many AdS black holes, for example RN AdS black hole \cite{25}.

\par
Figure 3 show that the isenthalpic and inversion curves with the fixed charge and accelerating parameter. One can see that the isenthalpic curve is divided into two regions by the inversion curve. The cooling region of a black hole is represented by the top of the inversion curve, whereas the lower part of the inversion curve represents the heating region of the black hole. The results shows that the cooling-heating phase transition of the black hole, and the inversion curve reflects the boundary of the phase transition.
\begin{center}
\includegraphics[width=7cm,height=6cm]{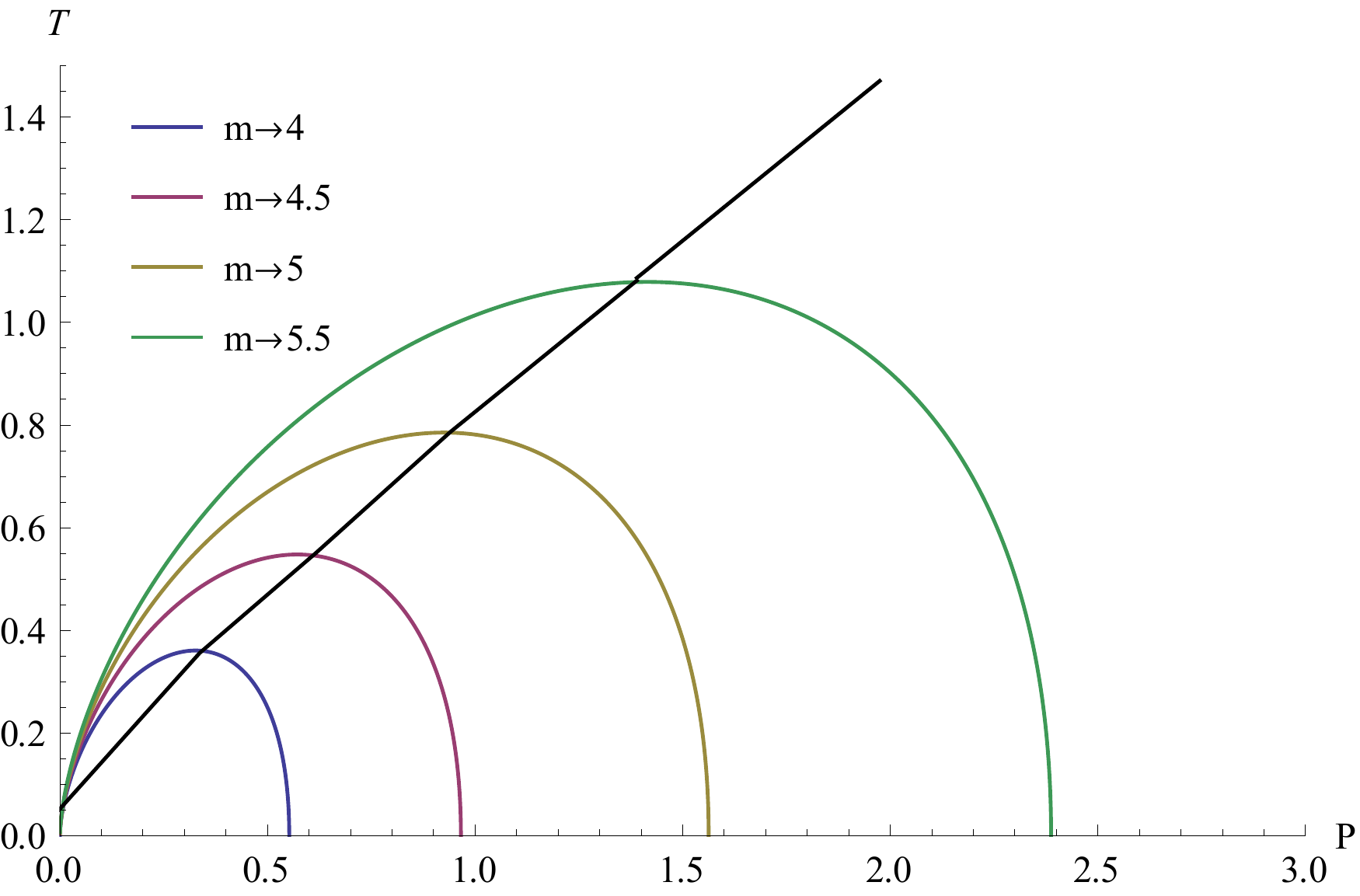}
\includegraphics[width=7cm,height=6cm]{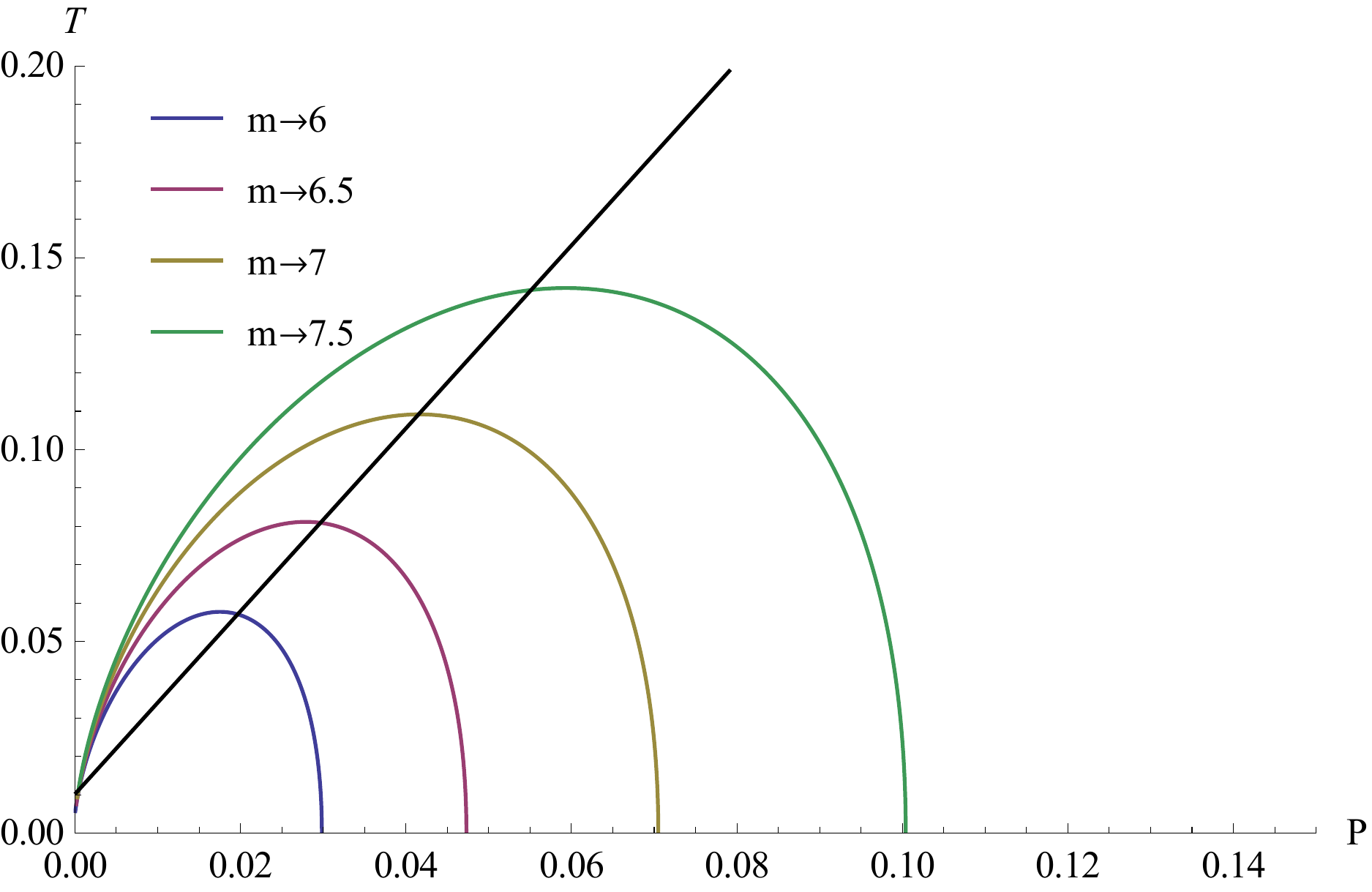}
\parbox[c]{15.0cm}{\footnotesize{\bf Fig~3.} 
Inversion and isenthalpic (constant mass) curves of the charged accelerating AdS black hole. {\em Panel (a)}-- black hole mass as $m=4,5.5,5,5.5$ and charge as $e=2$, {\em Panel (b)}-- black hole mass as $m=6,6.5,7,7.5$ and black hole charge as $e=4$. The accelerating parameter taking as $a=0.5$.}
\label{fig:3}
\end{center}

\par
Figure 4 further reflect that the isenthalpic curves under certain mass and charge to the influence of acceleration parameter $a$ on the cooling-heating phase transition. The isenthalpic curves shows a downward collapse trend with the increase of acceleration parameters $a$, meaning that the phase transition point of cooling-heating phase transition decreases gradually with the increase of acceleration constant $a$.
\begin{center}
\includegraphics[width=9cm,height=7cm]{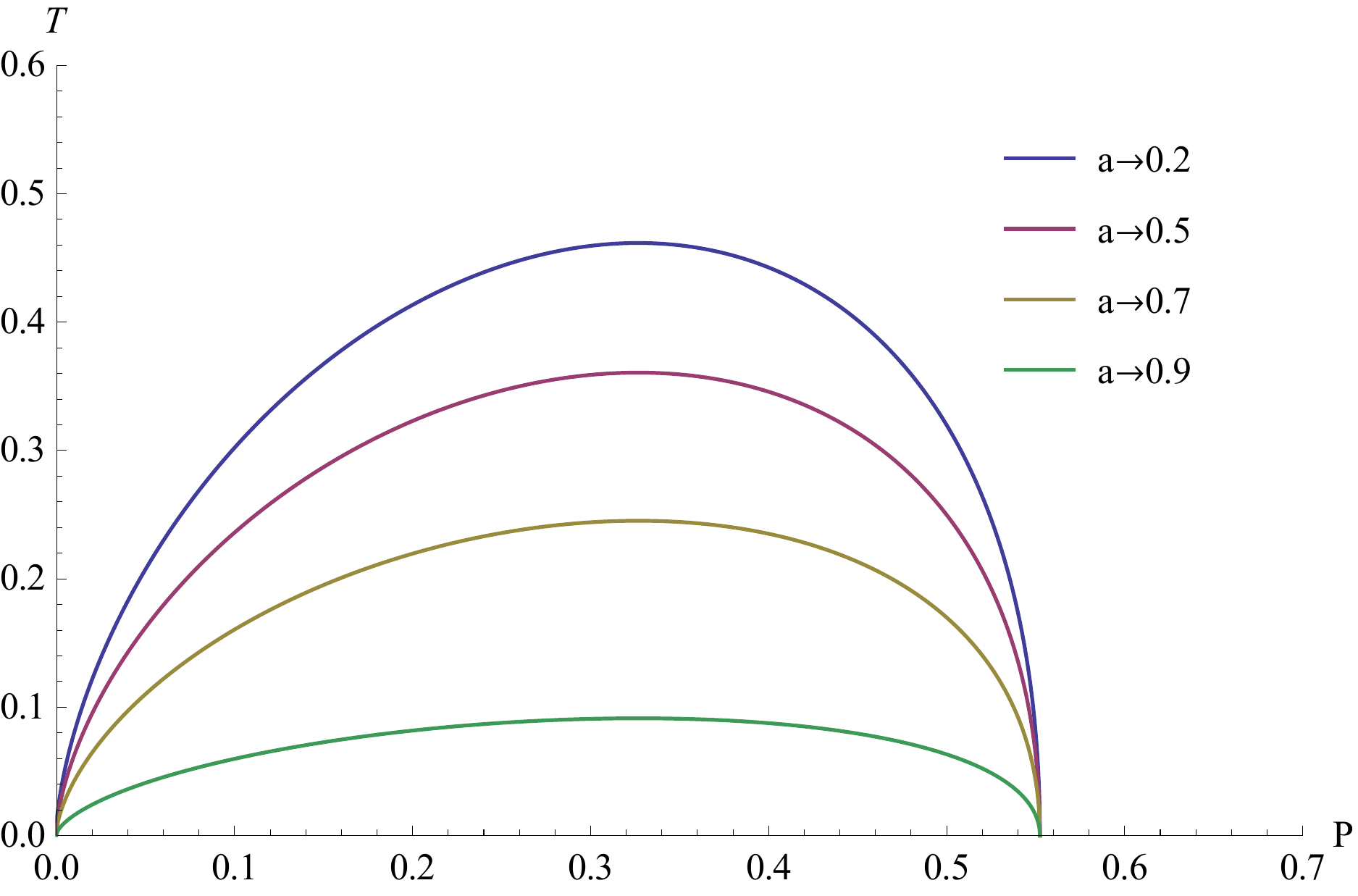}
\parbox[c]{15.0cm}{\footnotesize{\bf Fig~4.} 
The isenthalpic (constant mass) curves of the charged accelerating AdS black hole. From top to bottom, $a$=$0.2$, $0.5$, $0.7$, $0.9$. The mass as $m=4$ and the charge as $e=2$.}
\label{fig:4}
\end{center}

\section{Conclusions and Discussions}
\label{sec:4}
\par
The cooling-heating phase transition and critical behavior of the charged accelerating AdS black hole in extended phase space have been revealed in this analysis. The state equation and critical physics quantities of this black hole have been obtained. It is found that the universal ratio is different with the van der Waals system at critical point. Interestingly, the result returns to $3/8$ of the van der Waals system when the acceleration parameter $a$ is zero. Meanwhile, we also found that the critical exponents of this black is same with van der Waals system $(\alpha,\beta,\gamma,\delta)$ =(0,1/2,1,3), implying that this black hole phase transition can be similar to the van der Waals phase transition.

\par
Then, the cooling-heating phase transition of this black hole have been investigated. The highest point of the isenthalpic curves corresponds to the inversion point of temperature, it means that the black hole can be divided into cooling and heating two regions. We obtained the inversion temperature of the black hole and plotted the inversion curves, showing that the influence of charge $e$ and acceleration parameter $a$ on the cooling-heating phase transition. We found that the increase of $e$ leads to the increase of inversion temperature at the pressure constant, and the inversion temperature curve is lower for a larger $a$. This results indicate that the higher the temperature is needed to complete the cooling-heating phase transition for the larger charged the black hole. It is directly reflected that there is a positive correlation between the charge and the cooling heating phase transition, however, the acceleration factor shows the opposite state.

\par
Different from previous studies, the acceleration factor of our black hole promotes the occurrence of phase transition. A lower temperature to make the black hole cooling-heating phase transition under the same pressure. To further verify our results, the isenthalpic curves when both mass $m$ and charge $e$ are constants have been plotted in Figure 4. As a result, we found that the isenthalpic curves shows a downward collapse trend with the increase of acceleration parameters $a$, i.e. The phase transition point of cooling-heating phase transition decreases gradually with the increase of acceleration constant $a$.

\section*{Acknowledgments}
The authors would like to thank the anonymous reviewers for their helpful comments and suggestions, which helped to improve the quality of this paper. This work is supported by the National Natural Science Foundation of China (Grant No.11903025).

\clearpage

\end{CJK}
\end{document}